\input harvmac
\input epsf
\noblackbox
\newcount\figno
\figno=0
\def\fig#1#2#3{
\par\begingroup\parindent=0pt\leftskip=1cm\rightskip=1cm\parindent=0pt
\baselineskip=11pt
\global\advance\figno by 1
\midinsert
\epsfxsize=#3
\centerline{\epsfbox{#2}}
\vskip 12pt
\centerline{{\bf Figure \the\figno} #1}\par
\endinsert\endgroup\par}
\def\figlabel#1{\xdef#1{\the\figno}}
\def\pano{\par\noindent}

\def\pmb#1{\setbox0=\hbox{#1}%
 \kern-.025em\copy0\kern-\wd0
 \kern.05em\copy0\kern-\wd0
 \kern-.025em\raise.0433em\box0 }
\font\cmss=cmss10
\font\cmsss=cmss10 at 7pt
\def\half{{1\over 2}}
\def\rlx{\relax\leavevmode}
\def\Cop{\relax\,\hbox{$\kern-.3em{\rm C}$}}
\def\Rop{\relax{\rm I\kern-.18em R}}
\def\Nop{\relax{\rm I\kern-.18em N}}
\def\Pop{\relax{\rm I\kern-.18em P}}
\def\Mon{{\bf M}}
\def\Zop{\rlx\leavevmode\ifmmode\mathchoice{\hbox{\cmss Z\kern-.4em Z}}
 {\hbox{\cmss Z\kern-.4em Z}}{\lower.9pt\hbox{\cmsss Z\kern-.36em Z}}
 {\lower1.2pt\hbox{\cmsss Z\kern-.36em Z}}\else{\cmss Z\kern-.4em
 Z}\fi}


\def\H{{\cal H}}

\def\W{{\cal W}}

\def\ie{{\it i.e.}}

\lref\conslo{J.H.~Conway, N.J.A.~Sloane, {\it Sphere packings,
 lattices and groups}, 3rd ed.\ Springer (1999).}

\lref\god{P.~Goddard, {\it Meromorphic conformal field theory}, in:
`Infinite dimensional Lie algebras and Lie groups': Proceedings
of the CIRM Luminy Conference 1988, World Scientific (1989).}

\lref\dgmone{L.~Dolan, P.~Goddard, P.~Montague, {\it Conformal field
theory, triality and the Monster group}, Phys.\ Lett.\ {\bf B236}, 165
(1990).}

\lref\dgmtwo{L.~Dolan, P.~Goddard, P.~Montague, {\it Conformal field
theory of twisted vertex operators}, Nucl.\ Phys.\ {\bf B338}, 529
(1990).}

\lref\dgmthree{L.~Dolan, P.~Goddard, P.~Montague, {\it Conformal field
theories, representations and lattice constructions},
Commun.\ Math.\ Phys.\ {\bf 179}, 61 (1996).}

\lref\flm{I.~Frenkel, J.~Lepowsky, A.~Meurman, {\it Vertex operator
algebras and the Monster}, Academic Press (1988).}

\lref\dgh{L.~Dixon, P.~Ginsparg, J.A.~Harvey, {\it Beauty and the
beast: superconformal symmetry in a Monster module},
Commun.\ Math.\ Phys.\ {\bf 119}, 285 (1988).}

\lref\moon{J.H.~Conway, S.P.~Norton, {\it Monstrous moonshine},
Bull.\ London Math.\ Soc.\ {\bf 11}, 308 (1979).}

\lref\bor{R.E.~Borcherds, {\it Monstrous moonshine and monstrous Lie
superalgebras}, Invent.\ Math.\ {\bf 109}, 405 (1992).}

\lref\orbifold{L.~Dixon, J.~Harvey, C.~Vafa, E.~Witten, {\it Strings
on orbifolds I and II}, Nucl.\ Phys.\ {\bf B261}, 678 (1985) and
Nucl.\ Phys.\ {\bf B274}, 295 (1986).}

\lref\cardy{J.L.~Cardy, {\it Boundary conditions, fusion rules and
the Verlinde formula}, Nucl.\ Phys.\ {\bf B324}, 581 (1989).}

\lref\DLMimrn{C.~Dong, H.~Li, G.~Mason, {\it Compact automorphism
groups of vertex operator algebras}, International Mathematics
Research Notices {\bf 18}, 913 (1996).}

\lref\DLM{C.~Dong, H.~Li, G.~Mason, {\it Modular invariance of trace
functions in orbifold theory}, Commun.\ Math.\ Phys.\ {\bf 214}, 1
(2000); {\tt q-alg/9703016}.}

\lref\mckaystrauss{J.~McKay, H.~Strauss, {\it The $q$-series of
monstrous moonshine \& the decomposition of the head characters},
Commun.\ Alg.\ {\bf 18(1)}, 253 (1990).}

\lref\gannon{T.~Gannon, {\it Monstrous moonshine and the
classification of CFT}, {\tt math.QA/9906167}.}

\lref\gabreck{M.R.~Gaberdiel, A.~Recknagel, {\it Conformal boundary
states for free bosons and fermions}, J.\ High Energy Phys.\
{\bf 0111}, 016 (2001); {\tt hep-th/0108238}.}

\lref\janik{R.A.~Janik, {\it Exceptional boundary states at c=1},
Nucl.\ Phys.\ {\bf B618}, 675 (2001); {\tt hep-th/0109021}.}

\lref\kac{V.G.~Kac, {\it Vertex algebras for beginners},
Amer.\ Math.\ Soc.\ (1997).}

\lref\fhm{I.~Frenkel, Y.-Z.~Huang, J.~Lepowsky, {\it On axiomatic
approaches to vertex operator algebras and modules},
Mem.\ Am.\ Math.\ Soc.\ {\bf 104}, 1 (1993).}

\lref\IvanovTuite{R.I.~Ivanov, M.P.~Tuite,
{\it Rational generalised moonshine from abelian orbifoldings of the
moonshine module}, {\tt math.QA/0106027}.}

\lref\tensiondimension{J.A.~Harvey, S.~Kachru, G.~Moore,
E.~Silverstein, {\it Tension is dimension},
J.\ High Energy Phys.\ {\bf 0003}, 001 (2000);
{\tt hep-th/9909072}.}

\lref\GRW{M.R.~Gaberdiel, A.~Recknagel, G.M.T.~Watts,
{\it The conformal boundary states for SU(2) at level 1},
{\tt hep-th/0108102}, to appear in Nucl.\ Phys.\ {\bf B}.}

\lref\mrgrev{M.R.~Gaberdiel,
{\it Lectures on Non-BPS Dirichlet branes},
Class.\ Quant.\ Grav.\ {\bf 17}, 3483 (2000);
{\tt hep-th/0005029}.}

\lref\gabste{M.R.~Gaberdiel, B.~Stefa\'nski,
{\it Dirichlet branes on orbifolds},
Nucl.\ Phys.\ {\bf B578}, 58 (2000);
{\tt hep-th/9910109}.}

\lref\bgtwo{O.~Bergman, M.R.~Gaberdiel,
{\it Stable non-BPS D-particles},
Phys.\ Lett.\ {\bf B441}, 133 (1998);
{\tt hep-th/9806155}.}

\lref\diagom{D.-E.~Diaconescu, J.~Gomis,
{\it Fractional branes and boundary states in orbifold theories},
{\tt hep-th/9906242}.}

\lref\bcf{M.~Bill\'o, B.~Craps, F.~Roose,
{\it Orbifold boundary states from Cardy's condition},
J.\ High Energy Phys.\ {\bf 0101}, 038 (2001);
{\tt hep-th/0011060}.}

\lref\polchinski{J.~Polchinski,
{\it Dirichlet-branes and Ramond-Ramond charges},
Phys.\ Rev.\ Lett.\  {\bf 75}, 4724 (1995);
{\tt hep-th/9510017}.}

\lref\cg{B.~Craps, M.R.~Gaberdiel,
{\it Discrete torsion orbifolds and D-branes. II},
J.\ High Energy Phys.\ {\bf 0104}, 013 (2001);
{\tt hep-th/0101143}.}

\lref\fs{J.~Fuchs, Ch.~Schweigert,
{\it Symmetry breaking boundaries II. More structures; examples},
Nucl.\ Phys.\ {\bf B568}, 543 (2000);
{\tt hep-th/9908025}.}

\lref\atlas{J.H.~Conway, R.T.~Curtis, S.P.~Norton, R.A.~Parker,
R.A.~Wilson, {\it ATLAS of finite groups: maximal subgroups and
ordinary characters for simple groups}, Clarendon Press, Oxford
(1985).}

\lref\lew{D.C.~Lewellen, {\it Sewing constraints for conformal field
theories on surfaces with boundaries}, Nucl.\ Phys.\ {\bf B372}, 654
(1992).}

\lref\lauer{J.~Lauer, J.~Mas, H.P.~Nilles, {\it Twisted sector
representations of discrete background symmetries for two-dimensional
orbifolds}, Nucl.\ Phys.\ {\bf B351}, 353 (1991).}

\lref\brunner{I.~Brunner, R.~Entin, C.~R\"omelsberger,
{\it D-branes on T(4)/Z(2) and T-duality},
J.\ High Energy Phys.\ {\bf 9906}, 016 (1999);
{\tt hep-th/9905078}.}

\lref\gimpol{E.G.~Gimon, J.~Polchinski, {\it Consistency conditions
for orientifolds and D-manifolds}, Phys.\ Rev.\ {\bf D54}, 1667
(1996); {\tt hep-th/9601038}.}

\lref\dougmoore{M.R.~Douglas, G.~Moore, {\it D-branes, quivers and ALE
instantons}, {\tt hep-th/9603167}.}

\lref\mrgcorfu{M.R.~Gaberdiel, {\it  D-branes from conformal field
theory}, {\tt hep-th/0201113}.}

\lref\brr{I.~Brunner, A.~Rajaraman, M.~Rozali, {\it D-branes on
asymmetric orbifolds}, Nucl.\ Phys.\ {\bf B558}, 205 (1999);
{\tt hep-th/9905024}.}

\lref\kors{B.~Kors, {\it D-brane spectra of nonsupersymmetric,
asymmetric orbifolds and nonperturbative contributions to the
cosmological constant}, J.\ High Energy Phys.\ {\bf 9911}, 028 (1999);
{\tt hep-th/9907007}.}

\lref\proc{J.H.~Conway, {\it $Y_{555}$ and all that}, in:
`Groups, combinatorics \& geometry': Proceedings of the
LMS Durham Symposium 1990,  M.~Liebeck, J.~Saxl (eds.), Cambridge
University Press (1992).}

\lref\tuite{M.~Tuite, {\it On the relationship between monstrous
moonshine and the uniqueness of the moonshine module}, 
Commun. Math. Phys. {\bf 166} (1995) 495.}

\lref\cumminsgannon{C.J.~Cummins, T.~Gannon, {\it Modular equations
and the genus zero property of moonshine functions}, Invent.\ Math.\
{\bf 129} (1997) 413.}

\Title{\vbox{
\hbox{hep--th/0202074}
\hbox{EFI-02-63}
\hbox{KCL-MTH-02-05}}}
{\vbox{\centerline{Monstrous branes}
}}
\centerline{Ben Craps\footnote{$^\star$}{{\tt
e-mail: craps@theory.uchicago.edu}}$^{,a}$,
Matthias R.\ Gaberdiel\footnote{$^\dagger$}{{\tt
e-mail: mrg@mth.kcl.ac.uk}}$^{,b}$ and
Jeffrey A.\ Harvey\footnote{$^{\ddagger}$}{{\tt
e-mail: harvey@theory.uchicago.edu}}$^{,a,c}$}
\bigskip
\centerline{\it $^a$Enrico Fermi Institute, University of Chicago}
\centerline{\it 5640 S. Ellis Av., Chicago, IL 60637, USA}
\medskip
\centerline{\it $^b$Department of Mathematics, King's College London}
\centerline{\it Strand, London WC2R 2LS, U.K.}
\medskip
\centerline{\it $^c$Department of Physics, University of Chicago}
\centerline{\it 5640 S. Ellis Av., Chicago, IL 60637, USA}
\bigskip
\vskip1.5cm
\centerline{\bf Abstract}
\bigskip
\noindent
We study D-branes in the bosonic closed string theory whose
automorphism group is the Bimonster group (the wreath product of
the Monster simple group with $\Zop_2$). We give a complete
classification of D-branes preserving the chiral subalgebra of
Monster invariants and show that they transform in a
representation of the Bimonster. Our results apply more generally
to self-dual conformal field theories which admit the action of a
compact Lie group on both the left- and right-moving sectors.

\bigskip

\Date{February, 2002}

\newsec{Introduction}

The connection between the Monster sporadic group and modular
functions known as moonshine is one of the most peculiar and
mysterious facts in modern mathematics. Equally strange is the fact
that the construction of a Monster module which most naturally
encodes these connections uses the techniques and ideas of string
theory \flm.

There now exists a proof of the moonshine conjectures \bor, and steps have been
taken towards an explanation of their origin \refs{\tuite,\cumminsgannon}.
However, a fully satisfactory conceptual explanation of the connection 
between the Monster,
modular functions, and its appearance in string theory is still
lacking. It may be that some new physical ideas and techniques
will help to shed light on  the situation.

The current understanding of the Monster in string theory shows
that, in a particular closed string background, which will be
described later, the Bimonster acts as a symmetry group of the
perturbative spectrum of the string theory. (The Bimonster is the
wreath product of the Monster simple group with $\Zop_2$, \ie\ two
copies of the Monster group exchanged by an involution; see
subsection~2.2 for more details.) Over the last several years it
has been appreciated that string theory also contains
non-perturbative states whose mass scales like $1/g_s$ or
$1/g_s^2$ with $g_s$ the string coupling, and that these states
play a fundamental role in understanding the structure of string
theory \refs{\polchinski}. In the bosonic string theory in which the 
Bimonster appears, the best understood non-perturbative states are
Dirichlet branes or D-branes whose mass scales like $1/g_s$. The 
classification of D-brane states in various string backgrounds has
been an active area of research recently, and our aim here is to
use the techniques that have been developed to at least partially
classify the possible D-brane states in the bosonic string theory
with Bimonster symmetry. In doing so we will provide evidence that
the Bimonster extends from a symmetry of the perturbative spectrum
to a symmetry of the full spectrum of the theory.

The construction of the Monster module in \flm\ can be viewed, in
string theory language, as the construction of an asymmetric orbifold
of a special toroidal compactification of bosonic string theory. As a
result we will need to utilise a number of results regarding the
description of D-branes in orbifold conformal field theory. In the
next subsection we shall therefore review briefly, following
\refs{\cg}, some of the necessary background material.

\subsec{Orbifolds, D-branes and conformal field theory}

Let us begin by explaining the basic ideas underlying the orbifold
construction \refs{\orbifold}. Consider a closed string theory
compactified on a manifold ${\cal M}$ on which a group $\Gamma$ acts
as a group of symmetries. Roughly speaking, the orbifold by $\Gamma$
is the compactification on the quotient space  ${\cal M}/\Gamma$.  If
the action of the discrete group on ${\cal M}$ is not free, \ie\ if
${\cal M}$ has fixed points under the action of some elements in
$\Gamma$, then the resulting space is  singular. Despite such
classical singularities, string theory is however well-behaved on such
orbifolds.

More specifically we can describe the orbifold theory as
follows. Firstly, the theory consists of those states in the
original space of states $\H$ that are invariant under the action of
the orbifold group $\Gamma$. In addition, the theory has so-called
twisted sectors containing strings that are closed in
${\cal M}/\Gamma$ but not in ${\cal M}$. If the orbifold action has
fixed points, the twisted sector states describe degrees of freedom
that are localised at these fixed points; the presence of these
additional states is the essential reason for why string theory is
well-behaved despite these singularities.

The concept of an orbifold can be extended to a more general setting,
where neither the underlying conformal field theory $\H$ nor the
discrete symmetry group $\Gamma$ need to have a direct geometric
interpretation. (For example, in the case studied in this paper,
$\Gamma$ includes an asymmetric reflection, acting on the left-moving
string coordinates only.) In this context the
twisted sectors are then determined by the condition that the orbifold
conformal field theory should be modular invariant. If $\Gamma$ is
abelian one finds that there is one twisted sector $\H_h$ for each
element $h\in \Gamma$. Each twisted sector has to be projected again
onto the states that are invariant under the action of the orbifold
group $\Gamma$.
\smallskip

Next, we turn to the description of D-branes on orbifolds. Let us
first consider the case where the orbifold has a geometric
interpretation as ${\cal M}/\Gamma$. Then we can construct D-branes
as follows \refs{\gimpol,\dougmoore}: we consider a D-brane on the
covering space ${\cal M}$, and add to it images under the action of
$\Gamma$ so as to obtain a $\Gamma$-invariant configuration of
D-branes on ${\cal M}$. We then restrict the resulting open string
spectrum to those states that are invariant under the action of the
orbifold group. A typical orbifold-invariant configuration will
consist of $|\Gamma|$ D-branes on the covering space. The resulting
D-brane is then called a `bulk' brane, and it possesses moduli that
describe its position on ${\cal M}$.

On the other hand, if the original D-brane is localised at a singular
point of the orbifold, we need fewer preimages in the covering space
to make an orbifold invariant configuration; such branes are then
called `fractional' D-branes. Because they involve fewer preimage
branes, fractional D-branes cannot move off the singular point;
instead, a number of fractional D-branes have to come together in
order for the system to be able to move off into the bulk.
\smallskip

For orbifolds that do not have a simple geometric interpretation, it
is often useful to describe D-branes in terms of boundary states,
using (boundary) conformal field theory
methods.\footnote{$^\star$}{Actually, the conformal field theory
point of view is also powerful for geometric orbifolds; see for
example \refs{\bgtwo,\diagom,\gabste,\mrgrev,\bcf}.} D-branes can be
thought of as describing open string sectors that can be
added consistently to a given closed string theory. From the point of
view of conformal field theory, the construction of D-branes is
therefore simply the construction of permissible boundary
conditions. This problem has been studied for a number of years (for
a recent review see for example \refs{\mrgcorfu}).

In the conformal field theory approach, D-branes are described by
coherent `boundary states' that can be constructed in the
underlying (closed) string theory. These boundary states satisfy a
number of consistency conditions, the most important of which is
the so-called Cardy condition \refs{\cardy} (which we shall
analyse in detail in section~4). It arises from considering the
annulus diagram for which the two boundary conditions are
determined by two (possibly identical) D-branes, one for each
boundary. This diagram can be given two interpretations, depending
on which world-sheet coordinate is chosen as the world-sheet time.
From the closed string point of view the diagram describes the
tree-level exchange of closed string states between two sources
(D-branes). On the other hand, the diagram can also be interpreted
as a one-loop vacuum diagram of open strings with boundary
conditions described by the two D-branes. The requirement that
both the open and the closed string interpretations of the annulus
diagram should be sensible imposes strong restrictions on the
possible D-branes in a given closed string theory.

\subsec{Outline}

The paper is organised as follows. In section 2 we review the
construction of the Monster theory whose D-branes we want to
study. In section~3 we make use of the orbifold construction of the
Monster theory to anticipate the presence of certain bulk and
fractional branes in the D-brane spectrum. We then employ conformal
field theory techniques in section~4 in order to construct all
D-branes that preserve the chiral subalgebra of Monster invariants. We
demonstrate that the branes we construct satisfy all relative Cardy
conditions, and we show that they are complete in a suitable
sense. The D-branes are labelled by group elements in the Monster
group $\Mon$, and transform in the regular representation of both
copies of $\Mon$ (that defines a representation of the Bimonster). As
we shall explain, our analysis is actually valid for any self-dual
conformal field theory which admits the action of a compact Lie group
on both left- and right-moving sectors, and we therefore couch our
arguments in this more general setting. In section~5 we explain how
the `geometrical' D-branes of section~3 can be accounted for in terms
of the more abstract conformal field theory construction. Finally, we
end with some conclusions in section~6.

\newsec{The Monster theory}

\subsec{The Monster conformal field theory}

\noindent We are interested in the (bosonic) closed string theory
whose spectrum is described by the tensor product of two (chiral)
Monster conformal field theories,
\eqn\hfull{
\H = \H_M \otimes \overline{\H_M} \,,}
where $\H_M$ is the (chiral) Monster theory,
\eqn\hmonster{
\H_M = \H_{\Lambda_{L}} / \Zop_2 \,.}
Here $\Lambda_{L}$ is the Leech lattice, the (unique) even self-dual
Euclidean lattice of dimension $24$ that does not possess any points
of length square $2$ (see \refs{\conslo} for a good
explanation of  these matters), and $\H_{\Lambda}$ is the holomorphic
bosonic conformal field theory associated to the even, self-dual
lattice $\Lambda$. The lattice theory $\H_\Lambda$ is the conformal
field theory that consists of the states of the form
\eqn\lattice{
\prod_{i=1}^{n} \alpha^{j_i}_{-m_i} \, | p \rangle \,,}
where $j_i\in\{1,\ldots, d\}$, $m_1\geq m_2 \geq \cdots \geq m_n>0$ and
$p\in\Lambda$. Here $d$ is the dimension of the lattice which equals
the central charge of the conformal field theory $\H_\Lambda$,
$c=d$ with $d$ necessarily a multiple of 8,
and the oscillators $\alpha^i_n$, $i\in\{1,\ldots, d\}$,
$n\in\Zop$ satisfy the standard commutation relations
\eqn\osci{
[\alpha^i_m,\alpha^j_n] = m\, \delta^{ij}\, \delta_{m,-n} \,.}
More details about lattice conformal field theories can, for example,
be found in \refs{\god,\dgmthree}.

For the case under consideration $c=d=24$, and the $\Zop_2$ orbifold
acts on the $24$ oscillators, $\alpha^i_n$, $i=1,\ldots, 24$ by
\eqn\orbone{\alpha^i_n \mapsto - \alpha^i_n\,,}
and on the ground states $|p\rangle$ with $p\in\Lambda_{L}$ as
\eqn\orbtwo{ |p\rangle \mapsto |-p\rangle\,.}
The construction of this vertex operator algebra and the
demonstration that the Monster acts as its automorphism group
is due to Frenkel, Lepowsky and Meurman \refs{\flm}; the embedding of
this construction into conformal field theory has been discussed in
\refs{\dgh,\dgmone,\dgmtwo}.

The closed conformal field theory \hfull\ can be described as the
$\Zop_2\times\Zop_2$ orbifold of the compactification of the
closed bosonic string theory on the Leech torus, the quotient
space $\Rop^{24}/\Lambda_L$, with a special background $B$-field
turned on \refs{\dgh}. The two $\Zop_2$'s can be taken to be the
two asymmetric orbifolds that act on the left- and right-movers as
described above. Alternatively, we can also think of the theory as
an asymmetric $\Zop_2$ orbifold (where $\Zop_2$ acts as
above on the left-movers only) of the geometric $\Zop_2$ orbifold
of the Leech theory. Because of the background $B$-field we have just
mentioned, the generator of the asymmetric $\Zop_2$ actually differs
from the usual T-duality transformation that inverts all $24$
directions: indeed, T-duality not only reflects the left-movers, but
also rotates the right-movers by $D\equiv(G-B)^{-1}(G+B)$, where 
$G_{ij}=\delta_{ij}$ \refs{\lauer}. The distinction between the
usual T-duality and the asymmetric $\Zop_2$ we are considering here
will be important later on in section~3.

The untwisted sector of the chiral Monster theory \hmonster\
consists of those states of the lattice theory that are invariant
under the action of \orbone\ and \orbtwo. The twisted sector is
created by the action of half-integrally moded oscillators
$c^i_{r}$, $r\in\Zop+\half$, $i=1,\ldots,24$ satisfying
\eqn\oscitwis{ [c^i_r,c^j_s] = r\, \delta^{ij}\, \delta_{r,-s}\,,}
on the irreducible representation of the Clifford algebra
$\Gamma(\Lambda_L)$ associated to $\Lambda_L$.  This Clifford
algebra arises as a projective representation of $\Lambda_L/2
\Lambda_L$, generated by $\pm\gamma_i$, where $i=1,\ldots, 24$
labels a basis $k_i$ of $\Lambda_L$. The $\gamma_i$ satisfy the
relations
\eqn\clifford{ \gamma_{i} \gamma_{j} = (-1)^{k_i \cdot
k_j} \gamma_{j} \gamma_i\,,\qquad \gamma_i^2 = (-1)^{\half k_i^2}
\,.}
Since $k_i^2=4$ for $\Lambda_L$, each element $\gamma_i$
squares to one. The irreducible representation of this Clifford
algebra has dimension $2^{12}$, and thus the chiral theory has a
degeneracy of $2^{12}$ in the twisted sector. As before in the untwisted
sector, we also have to restrict the twisted sector states to be
$\Zop_2$-even, where the $\Zop_2$ generator acts on the oscillators  
as 
\eqn\orbc{c^i_r \mapsto - c^i_r\,,}
and on the degenerate twisted sector ground state $| \chi \rangle $ as
\eqn\orbctwo{| \chi \rangle  \mapsto
-| \chi \rangle.}
In the full theory \hfull\ we then have a degeneracy of
$2^{24}$ in the sector  where both left- and right-moving
oscillators are half-integrally moded; these correspond to
(certain linear combinations of) the $2^{24}$ fixed points under
the diagonal (geometrical) $\Zop_2$ orbifold. Two additional
sectors, where either the left- or the right-movers but not both
are half-integrally moded, each have a degeneracy of $2^{12}$;
these sectors are twisted by the $\Zop_2$ acting on the left- or
the right-movers, respectively.

\subsec{The Monster group and some subgroups}

\noindent The automorphism group of the chiral Monster conformal
field theory is the so-called Monster group \Mon, the largest
sporadic simple finite group. This is to say, for each $g\in\Mon$,
we have an automorphism of the conformal field theory,
\eqn\autom{g: \H_M \rightarrow \H_M \,,}
for which
\eqn\vertex{ g \,V(\psi,z)\, g^{-1} = V(g\, \psi,z) \,,}
where $V(\psi,z)$ is the vertex operator corresponding to the state
$\psi\in\H_M$. Furthermore,
\eqn\automone{ g\, |\Omega \rangle  =| \Omega \rangle
\qquad g \,| \omega \rangle  = g \,
L_{-2}\,| \Omega \rangle = L_{-2}\,| \Omega \rangle \,.}
Here $L_n$ denote the modes
of the Virasoro algebra,
\eqn\vira{ [L_m,L_n] = (m-n)\, L_{m+n} +
{c \over 12}\, m\, (m^2-1)\, \delta_{m,-n} \,,}
$|\Omega \rangle $ is the vacuum vector and
$|\omega \rangle =L_{-2}\,| \Omega \rangle$ is the conformal vector.
Since $|\omega \rangle$ is invariant under the action of $g\in\Mon$ it
follows from \vertex\ that
\eqn\act{ [g,L_n] = 0 \qquad \hbox{for all $g\in\Mon$,
$n\in\Zop$.}}
In particular, this implies that each eigenspace of
$L_0$ forms a representation of the Monster group, \Mon. The
theory has a single state with $h=0$, the vacuum $|\Omega\rangle$,
which transforms in the singlet representation of \Mon. There are no
states with $h=1$, and for $h=2$ we have $196 884$ states that
transform as
\eqn\monact{ {\bf 196 884} = {\bf 196 883} + {\bf 1}\,. }
Indeed, as we have seen, the state $L_{-2}\,|\Omega\rangle$ is
in the singlet representation of \Mon; the remaining states then
transform in the smallest non-trivial ($196 883$-dimensional)
representation of the Monster group. This pattern persists at
higher level \refs{\flm}.

The automorphism group of the full closed string theory \hfull\ is
then the so-called Bimonster group. The Bimonster is the wreath
product of the Monster group $\Mon$ with $\Zop_2$, \ie\ the
semi-direct product of $(\Mon \times \Mon)$ with $\Zop_2$ where
the generator $\sigma$ of $\Zop_2$ permutes the two copies of
$\Mon$. (A neat presentation of the Bimonster in terms of Coxeter
relations was conjectured by Conway, and subsequently proven by
Norton \refs{\proc}.) Indeed, the two copies of the Monster group
act on the left  and right chiral theory separately, and $\sigma$
is the symmetry that exchanges left- and right-movers (combined
with a shift of the $B$-field such that the background is
preserved).

The Monster group contains a subgroup $C$ whose action on the
Monster conformal field theory can be understood geometrically.
This subgroup is an extension of the simple Conway group $(\cdot
1) = (\cdot 0)/\Zop_2$ by an `extra-special' group denoted by
$2_+^{1+24}$; one therefore writes $C=2_+^{1+24}(\cdot 1)$. Since
$C$ will play a role in sections~3 and~5, we now review its
construction, following the exposition given in \refs{\dgmthree}.

\noindent The Conway group $(\cdot 0)$ is the group of automorphisms
of the Leech lattice,
\eqn\conzero{ (\cdot 0) = \hbox{Aut}(\Lambda_L) =
\Bigl\{ R \in \hbox{SO(24)} : R p \in \Lambda_L\; \hbox{for
$p\in\Lambda_L$} \Bigr\}\,.}
The centre of $(\cdot 0)$ contains one non-trivial element, the
reflection map $p\mapsto -p$ which we have used in the above orbifold
construction. Since this symmetry acts (by construction) trivially on
the orbifold theory, the automorphism group of the Monster conformal
field theory only involves the quotient group by this reflection
symmetry; this is the simple Conway group  $(\cdot 1)$.

\noindent Each element $R\in (\cdot 1)$ has a natural action on the
oscillators, given by
\eqn\cone{ \alpha^i_n \mapsto R_{ij}\,
\alpha^j_n\,, \qquad c^i_r \mapsto R_{ij}\, c^j_r\,,}
but the action
on the ground states is ambiguous. This ambiguity is responsible
for the extension of $(\cdot 1)$ mentioned above, as we now
describe. Let us extend the
$\gamma_i\equiv\gamma_{k_i}$ to being defined for arbitrary
$k\in\Lambda_L$, where the generators $\gamma_k$ now satisfy
\eqn\gam{ \gamma_k \gamma_l = (-1)^{k\cdot l} \gamma_l \gamma_k
\,, \qquad \gamma_{k} \gamma_l = \varepsilon(k,l) \gamma_{k+l}
\,,}
and $\varepsilon(k,l)$ are suitable signs. Because of the
first equation in \gam, these signs must satisfy
\eqn\epsprop{\varepsilon(k,l) = (-1)^{k\cdot l} \varepsilon(l,k)\,,}
while the associativity of the algebra product implies that
\eqn\epsproptwo{\varepsilon(k,l)\, \varepsilon(k+l,m)
     = \varepsilon(k,l+m)\, \varepsilon(l,m) \,.}
As we have explained before, the ground states of the twisted
sector form an irreducible representation of the algebra \gam.
Each element $R\in(\cdot 1)$ gives rise to an automorphism of the
gamma matrix algebra by
\eqn\Rauto{ \gamma'_k = \gamma_{Rk} \,,}
and this induces an automorphism of the corresponding
representation. Since all irreducible representations of the
Clifford algebra are isomorphic, there exists a unitary
transformation $S$ so that
\eqn\uni{ S\, \gamma_k \, S^{-1} = v_{R,S}(k)\, \gamma_{Rk} \,,}
where $v_{R,S}(k)=\pm 1$. This transforms $\varepsilon(k,l)$ into
\eqn\cobound{ \varepsilon'(k,l)
= {v_{R,S}(k+l) \over v_{R,S}(k)\, v_{R,S}(l)} \,
\varepsilon(k,l)\,.}
The action on the ground states is now defined by
\eqn\ground{ |p\rangle \mapsto v_{R,S}(p)\, |Rp\rangle
\qquad |\chi \rangle \mapsto S\,| \chi \rangle \,,}
where $|\chi \rangle$ denotes the $2^{12}$ ground states of the
twisted sector. This construction also applies to elements
$R \in (\cdot 0)$; for example, the generator of the $\Zop_2$ defined
by \orbone, \orbtwo, \orbc\ and \orbctwo\ corresponds to $R=-1$ and
$S=-1$ with $v_{R,S}(k)=1$.

\noindent The unitary transformation $S$ that enters in \ground\ is
only determined by \uni\ up to 
\eqn\freedom{ S\mapsto S\gamma\,,}
where $\gamma=\pm\gamma_l$ for some $l\in\Lambda_L$, \ie\
$\gamma\in\Gamma(\Lambda_L)$. For each $R\in(\cdot 1)$, there are
therefore $|\Gamma(\Lambda_L)|=2^{25}$ different choices for $S$,
thus leading to the extension of $(\cdot 1)$ by
$\Gamma(\Lambda_L)=2_+^{1+24}$. In particular, for $R=e$ and 
$S=\gamma_l$, 
%
\uni\ and \gam\ imply that 
\eqn\vel{v_{e,\gamma_l}(k)=(-1)^{k\cdot l}\,. }

The Monster group contains the involution $i$ which acts as $+1$ in
the untwisted sector and as $-1$ in the twisted sector. Actually the
centraliser of $i$ in {\bf M}, \ie\ the subgroup that consists of
those elements $g\in{\bf M}$ that commute with $i$, is precisely the
group $C=2_+^{1+24}(\cdot 1)$ that we have just described. In
particular, $i$ is therefore an element of $C$; it corresponds to
choosing $R=e$ in $(\cdot 1)$ and $S=-1$ in $\Gamma(\Lambda_L)$.

\subsec{Partition function and McKay-Thompson series}

\noindent The character or partition function of the (chiral)
Monster theory is given as
\eqn\part{\hbox{Tr}_{\H_M}(q^{L_0-{c\over 24}}) = j(\tau) - 744 \,,}
where $j(\tau)$ is the elliptic $j$-function,
\eqn\jfun{ j(\tau) =
{\Theta_{E_8}(\tau)^3 \over \eta(\tau)^{24}} = q^{-1} + 744 + 196
884\; q + 21 493 760\; q^2 + \cdots \,, \qquad
q=e^{2\pi i \tau}\,.}
Here $\Theta_{E_8}(\tau)$ is the theta function of the
$E_8$ root lattice,
\eqn\theta{ \Theta_\Lambda(\tau) =
\sum_{x\in\Lambda} q^{{1\over 2} x^2}\,,}
and $\eta(\tau)$ is the Dedekind eta-function,
\eqn\etafun{ \eta(\tau) = q^{{1\over 24}}
\prod_{n=1}^{\infty} (1-q^n) \,.}
The $j$-function (and therefore also \part) is a {\it modular}
function; this is to say, $j(\tau)$ is invariant under the action of
$SL(2,\Zop)$, \ie\
\eqn\jinv{ j\left( {a\tau + b \over c\tau + d} \right) = j(\tau) \,,
\qquad \pmatrix{a & b \cr c & d} \in SL(2,\Zop)\,.}

As we have explained above (see \act), the action of the Monster
group commutes with $L_0$. Thus it is natural to consider the
so-called McKay-Thompson series
\eqn\mt{ \chi_g(q) \equiv\hbox{Tr}_{\H_M}(g\, q^{L_0-{1}}) \,,}
for every element $g\in{\bf M}$. For $g=e$, the identity element of
the Monster group, \mt\ reduces to \part. Since the definition of \mt\
involves the trace over representations of ${\bf M}$, the
McKay-Thompson series only depends on the conjugacy class of $g$ in
{\bf M}. There are $194$ conjugacy classes, and the first fifty terms
in the power series expansions of \mt\ have been tabulated in
\refs{\mckaystrauss}.\footnote{$^\dagger$}{Since the coefficients
of $q^n$ in \mt\ are real (they are in fact all integers), the
McKay-Thompson series for $g$ and $g^{-1}$ agree; in fact, there
are 22 conjugacy classes with distinct inverses. Moreover, the two
distinct classes of order 27 turn out to have the same
McKay-Thompson series, so all in all there exist $194-22-1=171$
different McKay-Thompson series.}

The McKay-Thompson series have a number of remarkable
properties. In particular, for each $g\in{\bf M}$, $\chi_g(q)$ is a
Hauptmodule of a genus zero modular group. This is the key statement
of the moonshine conjecture of Conway and Norton \refs{\moon} that has
now been proven by Borcherds \refs{\bor}. (For a nice introduction to
`monstrous moonshine' see \refs{\gannon}.)

\newsec{D-branes of the Monster conformal field theory: some examples}

\noindent Our aim is to construct, and to some extent classify, the
D-branes of the Monster theory. In particular, we would like to
understand whether the D-brane states fall into representations of the
Bimonster group.

In the following we shall mainly concentrate on those D-branes
that preserve the subalgebra $\W$ of the full (chiral) Monster vertex
operator algebra that consists of the Monster invariant
states in $\H_M$. As we shall see, a complete classification for
these D-branes is possible, and one can show that they transform
in a representation of the Bimonster group. This result will emerge
as a special case of the much more general analysis in section 4.
That analysis combines well-known techniques from boundary
conformal field theory \refs{\cardy} with mathematical results on
vertex operator algebras \refs{\DLMimrn}. In the  present section,
we use the description of the Monster conformal field theory as an
orbifold of the Leech lattice to anticipate the presence of some of
these boundary states. In section 5 we shall then show how these
examples fit into the analysis of section 4.

\subsec{Fractional $D0-D24$ at the origin}

\noindent We are considering the $\Zop_2\times\Zop_2$ orbifold of the
Leech compactification. As we have mentioned in the paragraph
following  \orbtwo, we can think of this as an asymmetric $\Zop_2$
orbifold of the geometric $\Zop_2$ orbifold of the Leech theory. We
are interested in those configurations of D-branes on the geometric
$\Zop_2$ orbifold of the Leech theory that are invariant under the
asymmetric $\Zop_2$ (which differs from T-duality by a rotation of the
right-movers). The simplest such configuration is a fractional
D0-brane sitting at the origin together with a D24-brane without
Wilson lines. The corresponding D-brane in the orbifold theory is
described by a boundary state
\eqn\bsDzero{
{1\over\sqrt{2}}\Bigl(|\!|\hbox{D0}\rangle\!\rangle_U
+ |\!|\hbox{D0}\rangle\!\rangle_T
+|\!|\hbox{D24}\rangle\!\rangle_U
+ |\!|\hbox{D24}\rangle\!\rangle_T\Bigr)\,,}
where the subscripts $U$ and $T$ denote components in the untwisted
sector and in the sector twisted by the geometric $\Zop_2$,
respectively. Specifically, we have for the untwisted states
\eqn\utdefstwo{
|\!|\hbox{D0}\rangle\!\rangle_U =
\sum_{p} \exp\left(\sum_{n\geq 1}  {1\over n} \alpha^i_{-n}
\bar\alpha^i_{-n}\right)|(p,p)\rangle\, }
and
\eqn\utdefsone{
|\!|\hbox{D24}\rangle\!\rangle_U = \sum_{p}
\exp\left(\sum_{n\geq 1} - {1\over n}
\alpha^i_{-n} \bar\alpha^i_{-n}\right)|(p,-p)\rangle\,, }
with $|(p_L,p_R) \rangle$ denoting the momentum ground state in
${\cal H}$ with momentum $p_L,p_R \in \Lambda_L$.
By construction, this combination of
boundary states is invariant under the asymmetric $\Zop_2$. However,
our $\Zop_2$ does not simply correspond to the T-duality
transformation that inverts all $24$ directions. Indeed, because of
the background $B$-field mentioned after \orbtwo, the T-dual of
$|\!|\hbox{D0}\rangle\!\rangle_U$, would be the familiar
\eqn\dtwnfour{
|\!|\hbox{D24'}\rangle\!\rangle_U = \sum_{p}
\exp\left(\sum_{n\geq 1} - {1\over n}
\alpha^i_{-n} D_{ij}\,  \bar\alpha^j_{-n}\right)|(p,-p)\rangle\,,}
where $D_{ij}=(1-B)^{-1}_{il}(1+B)_{lj}$.
This expression can be obtained from \utdefsone\ by rotating
the right-movers with the matrix $D$. Physically, \utdefsone\
therefore describes a D24-brane with a (Born-Infeld) flux $F=-B$ which
compensates for the background $B$-field.

\noindent For the twisted  sector contribution we have similarly
\eqn\tdefstwo{
|\!|\hbox{D0}\rangle\!\rangle_T =
\exp\left(\sum_{r\geq 1/2}  {1\over r} c^i_{-r}
\bar c^i_{-r}\right)|x=0\rangle\, }
and%
\footnote{$^\ddagger$}{
In the twisted sector, the image of $|\!|\hbox{D0}\rangle\!\rangle_T$
under T-duality would involve all $2^{24}$ fixed points (see for
instance  \refs{\lauer, \brunner}).}
\eqn\tdefsone{
|\!|\hbox{D24}\rangle\!\rangle_T =
- \exp\left(\sum_{r\geq 1/2} - {1\over r}
c^i_{-r} \bar c^i_{-r}\right) |x=0\rangle\,, }
where $|x=0\rangle$ denotes the twisted sector ground state localised
at the fixed point $0$.

\subsec{More fractional branes}

Given this `geometric' D-brane, we can obtain a number of other
configurations with a clear geometric interpretation. First of
all, we can consider introducing Wilson lines for the D24-brane,
thereby placing the D0-brane on a different fixed point $y$ (where
$y\in\half\Lambda_L / \Lambda_L$).
In the untwisted sector of the above boundary states this corresponds
to introducing $p$-dependent signs in \utdefsone\ and \utdefstwo\ as
\eqn\utdefstwobis{
|\!|\hbox{D0},y\rangle\!\rangle_U =
\sum_{p}(-1)^{2\, y \cdot p}\, \exp\left(\sum_{n\geq 1}  {1\over n}
\alpha^i_{-n} \bar\alpha^i_{-n}\right)|(p,p)\rangle}
and
\eqn\utdefsonebis{
|\!|\hbox{D24},y\rangle\!\rangle_U = \sum_{p}(-1)^{2\, y\cdot p} \,
\exp\left(\sum_{n\geq 1} - {1\over n}
\alpha^i_{-n} \bar\alpha^i_{-n}\right)|(p,-p)\rangle \,.}
In the twisted sector the relevant modifications are
\eqn\tdefstwobis{
|\!|\hbox{D0},y\rangle\!\rangle_T =
\exp\left(\sum_{r\geq 1/2}  {1\over r} c^i_{-r}
\bar c^i_{-r}\right)|x=y\rangle}
and
\eqn\tdefsonebis{
|\!|\hbox{D24},y\rangle\!\rangle_T =
-\exp\left(\sum_{r\geq 1/2} - {1\over r}
c^i_{-r} \bar c^i_{-r}\right) |x=y\rangle\,.}

Since there are $2^{24}$ different fixed points, there are $2^{24}$ such
configurations. In addition, we
can also change the overall sign of the twisted sector
contributions; thus in total there should be $2^{25}$ such
configurations.

From section 2 it is clear that introducing these signs corresponds
precisely to the chiral action of
$\Gamma(\Lambda_L)=2_+^{1+24}$. Indeed, $i\in\Gamma(\Lambda_L)$
changes the overall sign of the twisted sector contributions, and the
other elements with $R=e$ in $(\cdot 1)$ introduce the correct
$p$-dependent signs in the untwisted sector (see \vel).

We can also apply a number of asymmetric reflections of the Leech lattice
to relate the D0-D24 combination to a Dp-D(24-p) combination. This can be
implemented by a suitable lift of an element in
$(\cdot 1)$ to the extra-special extension $C$ introduced before.
Combining these two constructions we therefore conclude that the image
of the above D24-D0 brane under the action of $C=2_+^{1+24}(\cdot 1)$
gives another geometric brane  configuration.

On the other hand, the other generators of the Monster group do
not map the D0-D24 brane into another geometrical configuration.
This is not really surprising since the oscillators $\alpha^i_n$
do not transform in a representation of the Monster group.
(After all, the smallest non-trivial representation of the Monster
group has dimension $196883$.) This is not in conflict with the
claim that the Monster group acts on the conformal field theory
since the modes $\alpha^i_n$ are {\it not} the modes of an actual
state of the theory --- they are the modes of the state
$\alpha^i_{-1}|\Omega\rangle$, but this state does not survive the
orbifold projection.

\subsec{Bulk branes}

\noindent Apart from the fractional D-brane states we have
constructed so far, one also expects there to be bulk D-branes,
which can move away from the fixed points of the orbifold. More
specifically, if we start with a bulk D0 brane at an arbitrary
point of the geometric $\Zop_2$ orbifold (\ie\ a pair of D0-branes at 
$x$ and $-x$ on the covering space), then adding two D24-branes
with Wilson lines $x$ and $-x$ (in suitable units) on the covering
space gives an orbifold invariant combination. This D-brane will
have moduli (corresponding to the position of one of the two
D0-branes, say), and so will be part of a continuum.

Such a bulk D-brane can be obtained by combining two coincident
fractional D-branes with cancelling twisted sector components. Indeed,
as we shall see later on, the open string spectrum between two such
states does indeed contain the appropriate marginal operators.

\newsec{Symmetric D-branes}

In principle, the only symmetry the boundary states of a (bosonic)
string theory are required to preserve is the conformal symmetry,
\ie\ the boundary states must satisfy
\eqn\introtwo{ \bigl(\, L_n
- \overline{L}_{-n}\,\bigr)\,|\!|\hbox{B}\rangle\!\rangle =0\,.}
In
general it is difficult to classify all such conformal boundary
conditions (see however \refs{\gabreck,\janik}), and one therefore
often restricts the problem further by demanding that the boundary
states preserve some larger symmetry. Examples are the familiar
Dirichlet or Neumann branes whose corresponding open strings
satisfy Dirichlet or Neumann boundary conditions at the ends. The
boundary states then preserve a U(1) current algebra for each
coordinate,
\eqn\introone{
\left(\alpha^i_{n} \pm \bar{\alpha}^i_{-n} \right) \,
|\!|\hbox{B}\rangle\!\rangle=0\,,}
where $\alpha^i_n$ are the modes associated to $X^i$, and the sign
determines whether $X^i$ obeys a Dirichlet or a Neumann condition
on the world-sheet boundary. Typically, the more symmetries one
requires a D-brane to preserve, the easier it is to construct and
classify the relevant boundary states. However, in general one
then only finds a subset of all physically relevant boundary
states.

For the case of the Monster theory we have so far 
only considered D-branes that consist of orbifold invariant
combinations of Dirichlet and Neumann branes. While these are
consistent D-branes, they are unlikely to describe {\it all} the
lightest D-branes of the theory since none of these branes couples to
the asymmetrically twisted closed string states. Furthermore, it  
is rather unnatural to consider gluing conditions that involve the
modes $\alpha^i_n$, since these modes are not actually present in the 
theory. (The modes are only present in the theory before orbifolding,
and thus the characterisation of these branes relies on a specific
realisation of the Monster conformal field theory in terms of an
orbifold construction.) 

Instead, we will in the following consider D-branes that are
characterised by a gluing condition that can be formulated within the
Monster conformal field theory. More specifically, we shall analyse
the branes that preserve the subalgebra of the original vertex
operator algebra consisting of Monster-invariant states. We will
denote this $W$-algebra by ${\W}$; it contains the Virasoro algebra,
but is in fact strictly larger. Let $W_n$ denote the modes of an
element of $\W$. Then the gluing condition we will impose,
generalising \introtwo, is
\eqn\wglue{
\Bigl(W_n - (-1)^s \bar W_{-n}\Bigr)\,
|\!|\hbox{B}\rangle\!\rangle =0\,,}
where $s$ is the spin of $W$.

As we shall see, this gluing condition is
restrictive enough to allow a complete classification of those
solutions. In section~5 we will discuss which of the examples in
section~3 are captured by this construction.

It turns out that the mathematical results we need in our
construction and classification of these symmetric boundary states
are known in much wider generality. Therefore, in the present
section, we will work in a broader framework than strictly
necessary to analyse the Monster theory.

\subsec{General framework}

We will work in the general framework studied in \refs{\DLMimrn}.
Suppose $\H_0$ is a simple vertex operator algebra, \ie\ a vertex
operator algebra that does not have any non-trivial ideals. (An
introduction to these matters can be found in
\refs{\kac,\flm,\fhm}.) Let us furthermore assume that $\H_0$
admits a continuous  action of a compact Lie group $G$ (which may
be finite). In the example of the Monster theory, $\H_0$
corresponds to the chiral Monster theory $\H_M$ and $G$ is the
Monster group \Mon. Let $\W$ be the vertex operator subalgebra of
$\H_0$ consisting of the $G$-invariants. 

\noindent The main result of \refs{\DLMimrn} is that $\H_0$ can be
decomposed as
\eqn\decom{ {\cal H}_0 = \bigoplus_{\lambda} R_\lambda\otimes
\H_\lambda \,,}
where the sum runs over all irreducible representations
$R_\lambda$ of $G$, and each $\H_\lambda$ is an irreducible
representation of $\W$.\footnote{$^\star$}{For the case of the
Monster theory, this result implies that $\W$ must be strictly larger
than the Virasoro algebra: otherwise the modular invariant partition
function \part\ would equal a finite sum of irreducible Virasoro
characters with $c=24$. One can also check directly, by comparing
characters, that $\W$ contains at least an additional primary field of
conformal weight $12$.}
Moreover, the $\H_\lambda$ are inequivalent for different
$\lambda$. The first few terms of the characters of $\H_\lambda$ can
be found in \refs{\mckaystrauss}.\footnote{$^\dagger$}{Table~2 of
that paper does not contain any entries for the (conjugate pairs of)
irreducible Monster representations IRR16-IRR17 and IRR26-IRR27. The
corresponding $\H_\lambda$ are not trivial, but they only contain
states whose conformal weight is bigger than $h=51$.}

\noindent The total space of states  (the generalisation of $\H$, see
\hfull) then has the decomposition
\eqn\fulldecom{ \H_0 \otimes \overline {\H_0}  =
\bigoplus_{\lambda,\mu}\,
\Bigl(R_\lambda\otimes \overline{R_\mu}\Bigr)
\otimes \Bigl(\H_\lambda\otimes \overline{\H_\mu}\Bigr) \,,}
where $\overline{R}$ denotes the conjugate \Mon-representation of $R$,
and the sum extends over all tensor products of irreducible
representations of $G$.

In the following we shall construct boundary states that preserve
$\W$ for this theory. We shall make use of the fact that 
$\H_0 \otimes \H_0$ has the decomposition \fulldecom. In general,
$\H_0 \otimes \H_0$ is only the vacuum sector of the full conformal 
field theory, and the theory also contains sectors that correspond to
other representations of $\H_0$. A priori, we do not know whether
these other sectors also have a decomposition as \fulldecom, and
we shall therefore restrict ourselves to {\it self-dual} theories,
\ie\ theories for which the full conformal field theory is
actually given by the vacuum sector alone. This is clearly the
case for the Monster theory. Self-dual conformal field theories
have the property that their character is invariant under the
S-modular transformation.

\subsec{Ishibashi states}

We are interested in constructing D-branes that preserve the full
$\W$-symmetry. Each such D-brane state can be written in terms of
$\W$-Ishibashi states; the Ishibashi state is uniquely fixed (up
to normalisation) by the gluing condition \wglue, and for each
term in \fulldecom\ for which the left- and  right-moving
$\W$-representations are conjugate, we can construct one such
Ishibashi state. (See for example \refs{\mrgcorfu} for a review of
these issues.) Thus we see from \fulldecom\ that the
$\W$-Ishibashi states are labelled just like matrix elements of
representations of $G$,
\eqn\ishi{ |\,R_\lambda ; i,\bar \jmath \,\rangle\!\rangle \in
\Bigl(R_\lambda\otimes \overline{R_\lambda}\Bigr)
\otimes
\Bigl(\H_\lambda\otimes \overline{\H_\lambda}\Bigr) \,,}
where $i\in R_\lambda$, $\bar \jmath \in\overline{R_\lambda}$ are a
basis for the representation $R_\lambda$ and $\overline{R_\lambda}$,
respectively. The relevant overlap between these Ishibashi states is
given as
\eqn\overlap{ \langle\!\langle R_\lambda;i_1,\bar \jmath_1 |\,
q^{\half(L_0+\bar{L}_0-{c\over12})}\, |\, R_\mu; i_2,\bar \jmath_2\,
\rangle\!\rangle = \delta_{\lambda,\mu}\, \delta_{i_1,i_2}
\,\delta_{\bar \jmath_1,\bar \jmath_2}\, \chi_{\H_\lambda}(q)\,,}
where $\chi_{\H_\lambda}(q)$ is the character of the irreducible
$\W$-representation $\H_\lambda$.

\subsec{Consistent boundary states}

Next we want to construct actual D-brane states, which are certain
linear combinations of the Ishibashi states. D-brane states are
(at least partially) characterised by the property that they
satisfy Cardy's condition \refs{\cardy}, \ie\ that they give rise
to a positive integer number of $\W$-representations (or more
generally, Virasoro representations) in the open string. One
D-brane state that satisfies this condition can be easily
constructed: it is given by
\eqn\ident{ |\!| e \rangle\!\rangle =
\sum_{\lambda;i} |\, R_\lambda; i,\bar \imath\, \rangle\!\rangle \,,}
where
the sum extends over all irreducible representations $R_\lambda$ of
$G$, and $i$ labels a basis of the representation $R_\lambda$. In
order to check that $|\!| e \rangle\!\rangle$ satisfies the Cardy 
condition, we observe that
\eqn\calcone{ \langle\!\langle e |\!|\,
q^{\half(L_0+\bar{L}_0-{c\over12})}\, |\!| e \rangle\!\rangle =
\sum_{\lambda;i} \chi_{\H_\lambda}(q) = \sum_{\lambda}
\dim(R_\lambda) \, \chi_{\H_\lambda}(q) \equiv F(\tau)\,,}
where we have used that $\H_0$ decomposes as in \decom, and where
$F(\tau)$ is the character (or partition function) of the chiral
conformal field theory $\H_0$. The character of a self-dual theory is
invariant under the modular transformation $\tau\mapsto -1/\tau$, and
therefore 
\eqn\calctwo{ \langle\!\langle e |\!|\,
q^{\half(L_0+\bar{L}_0-{c\over12})}\, |\!| e \rangle\!\rangle =
F(-1/\tau)=\sum_{\lambda} \dim(R_\lambda) \,
\chi_{\H_\lambda}(\tilde{q})\,,}
where
$\tilde{q}=e^{-2\pi i/\tau}$. Since $\dim(R_\lambda)$ are positive
integers, this demonstrates that the boundary state $|\!| e
\rangle\!\rangle$ satisfies the Cardy condition. For the special
case of the Monster theory, $F(\tau)=j(\tau)-744$, which is indeed
invariant under the S-modular transformation.
\medskip

As our notation suggests the boundary state \ident\ is
associated to the identity element of the automorphism group $G$. We
want to show next that there is actually a boundary state for each
group element of $G$. The different boundary states are transformed
into one another by the left-action of $G$. Thus we define 
\eqn\genbrane{ |\!| g \rangle\!\rangle = g \, |\!| e
\rangle\!\rangle = \sum_{\lambda;i, j} D^{R_\lambda}_{ji}(g)
\, |\, R_\lambda;  j,\bar\imath\, \rangle\!\rangle \,,}
where $D^{R}_{ji}(g)$ is the matrix element of $g\in G$
in the representation $R$,
\eqn\matrixel{ \sum_j
D^{R}_{lj}(h) \, D^{R}_{ji}(g) = D^{R}_{li}(hg)\,.}
The self-overlap of each of these
branes is in fact the same as \calcone\ above: it follows directly
from the definition of \genbrane\ that
\eqn\calcthree{
\langle\!\langle g |\!|\, q^{\half(L_0+\bar{L}_0-{c\over12})}\,
|\!| g \rangle\!\rangle = \sum_{\lambda;i,j}
\overline{D^{R_\lambda}_{ji}(g)} \, D^{R_\lambda}_{ji}(g) \,
\chi_{\H_\lambda}(q)\,.}
Since each group representation can be taken to be unitary, we have
\eqn\calcfour{ \sum_{i,j}
\overline{D^{R_\lambda}_{ji}(g)} \, D^{R_\lambda}_{ji}(g) =
\sum_{i,j} D^{R_\lambda}_{ij}(g^{-1}) \, D^{R_\lambda}_{ji}(g) =
\sum_i D^{R_\lambda}_{ii}(e) = \dim(R_\lambda) \,.} Inserting
\calcfour\ into \calcthree\ we thus reproduce \calcone. Incidentally,
this also shows that all these D-branes have the same mass, since
the mass is determined by the $q \rightarrow 0$ limit of
\calcthree\ \refs{\polchinski,\tensiondimension}.

It remains to show that the overlap between two different branes
of the form \genbrane\ also gives rise to a positive integer
number of representations of $\W$ in the open string. Using the
same argument as above in \calcthree\ and \calcfour\ we now find
\eqn\calcfive{ \eqalign{\langle\!\langle g |\!|\,
q^{\half(L_0+\bar{L}_0-{c\over12})}\, |\!| h \rangle\!\rangle & =
\sum_{\lambda;i,j} \overline{D^{R_\lambda}_{ji}(g)} \,
D^{R_\lambda}_{ji}(h) \, \chi_{\H_\lambda}(q)\cr & =
\sum_{\lambda} \hbox{Tr}_{R_\lambda}(g^{-1} h) \,
\chi_{\H_\lambda}(q) \cr & = \hbox{Tr}_{\H_0} (g^{-1} h \,
q^{L_0-{c\over 24}}) \,.}}
It follows from standard orbifold considerations \refs{\orbifold} that
under $\tau\mapsto -1/\tau$ we have
\eqn\orbi{ \hbox{Tr}_{\H_0} (\hat{g} \,q^{L_0-{c\over 24}}) =
\hbox{Tr}_{\H^{\hat{g}}} (\tilde{q}^{L_0-{c\over 24}}) \,,}
where we have written $\hat{g}=g^{-1}h$, and $\H^{\hat{g}}$ is the
(unique) $\hat{g}$-twisted representation of the conformal field
theory $\H_0$ \refs{\DLM}. Since the $\hat{g}$-twist acts trivially on
the generators of $\W$, we can decompose the representation
$\H^{\hat{g}}$ in terms of representations of $\W$ as
\eqn\decomtwis{ \H^{\hat{g}} = \bigoplus_{j} D_j \otimes \H_j \,,}
where each $\H_j$ is an irreducible representation of $\W$, and
$D_j$ is some multiplicity space. Thus it follows that
\eqn\calcseven{\langle\!\langle g |\!|\,
q^{\half(L_0+\bar{L}_0-{c\over12})}\, |\!| h\rangle\!\rangle =
\sum_j \dim(D_j) \,\chi_{\H_{j}}(\tilde{q})\,.}
In particular, this therefore implies that the relative overlaps also
satisfy Cardy's condition.

For the case of the Monster theory, the last line of \calcfive\ is
precisely the McKay-Thompson series \mt, which thus appears very
naturally in the study of monstrous D-branes!

\subsec{Completeness}

The construction of a set of consistent boundary states in the
previous subsection was fairly general. Now we want to argue that
at least in some cases (including the Monster theory) this set is
complete, in the sense that it contains all (fundamental) D-branes
preserving $\W$. In the following we shall restrict ourselves to the
case where $G$ is a finite group.

If $G$ is a finite group, there are only finitely many $\W$-preserving
Ishibashi states. Whenever this is the case, one can show the
completeness of the boundary states by the following algebraic
argument: suppose there are $N$ $\W$-Ishibashi states,
$|\hbox{I}_1\rangle\!\rangle,\ldots,|\hbox{I}_N\rangle\!\rangle$, and
that we have managed to find $N$ boundary states,
$|\!|\hbox{B}_1\rangle\!\rangle, \ldots,
|\!|\hbox{B}_N\rangle\!\rangle$ that are linearly independent over
the complex numbers. (This is the case in our example as we shall
show momentarily.) Then since every boundary state is a linear
combination of Ishibashi states, we can express the $N$ Ishibashi
states in terms of the $N$ boundary states, \ie\ we can find an
invertible matrix $A$ such that
\eqn\rel{ |\hbox{I}_j\rangle\!\rangle = \sum_{i} A_{ij}\,
|\!|\hbox{B}_i\rangle\!\rangle\,.}
Suppose now that there exists another boundary state
$|\!|\hbox{B}\rangle\!\rangle$ that is compatible with the
boundary states $|\!|\hbox{B}_i\rangle\!\rangle$ with $i=1,\ldots,N$.
(By this we mean that the various overlaps between
$|\!|\hbox{B}\rangle\!\rangle$ and the
$|\!|\hbox{B}_i\rangle\!\rangle$ lead to
positive integer numbers of $\W$ characters in the open string
channel.) Since $|\!|\hbox{B}\rangle\!\rangle$ is a boundary state, it
can be written as a linear combination of Ishibashi states, and
therefore, because of \rel, as a linear combination of the
boundary states $|\!|\hbox{B}_i\rangle\!\rangle$. Thus we have shown
that
\eqn\expnd{ |\!|\hbox{B}\rangle\!\rangle = \sum_{i} C_i\,
|\!|\hbox{B}_i\rangle\!\rangle\,,}
where the $C_i$ are some (in general complex) constants. In order to
prove that the $|\!|\hbox{B}_i\rangle\!\rangle$ are all the
(fundamental) boundary states it therefore only remains to show that
the $C_i$ are in fact non-negative integers. This will typically
follow from the fact that $|\!|\hbox{B}\rangle\!\rangle$ is compatible
with the $|\!|\hbox{B}_i\rangle\!\rangle$ in the sense described
above. For example, if the $|\!|\hbox{B}_i\rangle\!\rangle$ have the
property that the overlap between $|\!|\hbox{B}_i\rangle\!\rangle$ and
$|\!|\hbox{B}_j\rangle\!\rangle$ only contains the vacuum
representation in the open string provided that $i=j$, and that the
vacuum representation occurs with multiplicity one if $i=j$ (again
this is the case for the Monster theory as we shall show momentarily)
then this can be shown as follows. We consider the overlap
\eqn\overlap{\langle\!\langle
\hbox{B}_i |\!|\, q^{\half(L_0+\bar{L}_0-{c\over12})}\,|\!|\,
\hbox{B}\rangle\!\rangle\,,}
and transform into the open string description. From the above
assumption and \expnd\ it then follows that the vacuum character in
the open string occurs with multiplicity $C_i$. Thus it follows that
$C_i$ has to be a non-negative integer since
$|\!|\hbox{B}\rangle\!\rangle$ is compatible with
$|\!|\hbox{B}_i\rangle\!\rangle$.

For the case at hand, one can actually show that the two
assumptions made above are satisfied. First of all, it follows
from the above analysis that there are
\eqn\noish{ \sum_{\lambda} \dim(R_\lambda)^2 = \dim(G)}
Ishibashi states, which therefore agrees with the number of boundary
states described by \genbrane. By the Peter-Weyl Theorem (or the
appropriate simpler statement for finite groups) these $\dim(G)$
boundary states are linearly independent. Given the above argument,
this shows that the boundary states are all fundamental boundary
states provided they are `orthogonal', \ie\ provided that the identity
only arises in the open string of the overlap of each boundary
state with itself (where it arises with multiplicity one). For the
Monster theory, the latter statement is obviously correct since
the open string overlap between each boundary state and itself is
simply $j(\tilde{q})-744$ which starts indeed with
$1\tilde{q}^{-1} + \cdots$. Thus it only remains to check that the
overlap between different boundary states starts with
$0\tilde{q}^{-1} + \cdots$ in the open string. This is simply the
question of what the leading behaviour of the S-modular transform
of the different McKay-Thompson series (for $g\ne e$) is. It has
recently been argued that the only McKay-Thompson series that has a
term of order $\tilde{q}^{-1}$ in its S-modular transform is the
series associated to the identity element \refs{\IvanovTuite}. We have
also checked this property for a number of McKay-Thompson series
explicitly.
\medskip

We have thus shown that there are precisely $|\Mon|$ $\W$-preserving
boundary states $|\!| g \rangle\!\rangle$, labeled by $g\in\Mon$. It
is interesting to ask how these boundary states transform under the
Bimonster group. First of all, it is easy to verify that the elements
of $\Mon\times\Mon$ act as
\eqn\leftright{
(h_L,h_R) \, |\!| g \rangle\!\rangle
= |\!| h_L\, g\, h_R^{-1} \rangle\!\rangle\,.
}
Indeed, we calculate
\eqn\haction{
\eqalign{
(h_L,h_R) \, |\!| g \rangle\!\rangle & = (h_L,h_R)\,
\sum_{\lambda;i,j}
D^{R_\lambda}_{ji}(g)\, |\, R; j,\bar\imath\, \rangle\!\rangle \cr
& = \sum_{\lambda;i,j,k,l}
D^{R_\lambda}_{ji}(g)\; D^{R_\lambda}_{kj}(h_L)\;
\overline{D^{R_\lambda}_{li}(h_R)}\;
|\, R_\lambda; k,\bar{l}\, \rangle\!\rangle \cr
& = \sum_{\lambda;i,j,k,l}
D^{R_\lambda}_{kj}(h_L)\; D^{R_\lambda}_{ji}(g)\;
D^{R_\lambda}_{il}(h_R^{-1}) \;
|\, R_\lambda; k,\bar{l}\, \rangle\!\rangle \cr
& = \sum_{\lambda;k,l}
D^{R_\lambda}_{kl}(h_L\,g\, h_R^{-1})\;
|\, R_\lambda; k,\bar{l}\, \rangle\!\rangle \cr
& = |\!| h_L\, g\, h_R^{-1} \rangle\!\rangle\,.}
}
Next we note that the generator $\sigma$ of the $\Zop_2$ that
exchanges the left- and right-movers is an anti-linear map that
replaces the Ishibashi states 
$|\, R_\lambda; j,\bar \imath\, \rangle\!\rangle$ by
$|\, R_\lambda; i,\bar \jmath\, \rangle\!\rangle$. It therefore acts
on the boundary states as
\eqn\twistact{\eqalign{
\sigma \, |\!| g \rangle\!\rangle & =
\sum_{\lambda;\bar\jmath, i} \overline{D^{R_\lambda}_{ji}(g)}
\, |\, R_\lambda;  i,\bar\jmath\, \rangle\!\rangle \cr
& = \sum_{\lambda;\bar\jmath, i} D^{R_\lambda}_{ij}(g^{-1})
\, |\, R_\lambda;  i,\bar\jmath\, \rangle\!\rangle
= |\!| g^{-1} \rangle\!\rangle\,,}}
where we have again used that the representations of the Monster group
are unitary. The actions \leftright\ and \twistact\ combine to give a
full representation of the Bimonster group since
\eqn\bimonster{
\sigma \, (h_1,h_2) \, |\!| g \rangle\!\rangle
= |\!| h_2\, g^{-1}\, h_1^{-1} \rangle\!\rangle =
(h_2,h_1)\, \sigma \, \, |\!| g \rangle\!\rangle  \,.}
Thus we have shown that the $\W$-preserving boundary states fall into
a representation of the Bimonster group.

\subsec{Factorisation constraint}

In the previous subsections we have constructed a family of
$\W$-preserving boundary states that satisfy all relative Cardy
conditions. Furthermore, we have shown that this set of boundary
states is complete. In addition to the Cardy conditions, consistent
boundary states also have to satisfy the `sewing relations' of
\refs{\lew}. One of these conditions is the factorisation (or cluster)
condition that requires that certain bulk-boundary structure constants
satisfy a set of non-linear equations (sometimes also referred to as
the classifying algebra in this context). It was shown in \refs{\fs}
that a $\W$-preserving boundary state $|\!|\hbox{B}\rangle\!\rangle$
satisfies this factorisation constraint provided that it preserves the
full symmetry algebra $\H_0$ up to conjugation by an element in
$g\in G$, \ie\ provided
\eqn\factori{
\Bigl(g\, S_n\, g^{-1} - (-1)^s \bar S_{-n}
\Bigr)\, |\!|\hbox{B}\rangle\!\rangle =0\,,}
for all modes of fields in $\H_0$. (Here $g\in G$ depends on the
boundary condition $|\!|\hbox{B}\rangle\!\rangle$.) Since $W_n\in\W$
is invariant under the action of $g\in G$, \factori\ contains \wglue\
as a special case.

As we shall now explain, the boundary states we have
constructed actually satisfy \factori; in fact, we have
\eqn\factsol{
\Bigl(g\, S_n\, g^{-1} - (-1)^s \bar S_{-n}
\Bigr)\, |\!|g\rangle\!\rangle =0}
for each $g\in\Mon$. Given the decomposition \fulldecom,  the boundary
state corresponding to $|\!|e\rangle\!\rangle$ is the unique boundary
state that preserves the full symmetry algebra (the theory only
contains a single Ishibashi state that preserves this algebra), and
thus \factsol\ holds for $g=e$. The general statement then follows
from this using \leftright.

\newsec{Application to the Monster: fractional and bulk branes}

Let us now return to the specific case of the Monster theory. As
we have shown in the previous section, the D-branes that preserve
the $W$-algebra of Monster invariants $\W$ are labelled by group
elements in $\Mon$. We want to analyse now how the various D-branes
that we constructed in section~3 fit into this analysis. In order to
do so it is useful to describe the `geometrical' boundary states of
section~3.1 in more detail.

\subsec{Fractional D0-D24 at the origin}

In the untwisted sector of the geometric orbifold, the constituent
boundary states are (up to normalisation) given as
\eqn\dtf{
|\!|\hbox{D24}\rangle\!\rangle_U =
\sum_{p} \exp \left(\sum_{n\geq 1} - {1\over n}
\alpha^i_{-n} \bar\alpha^i_{-n}\right)|(p,-p)\rangle}
and
\eqn\dzero{
|\!|\hbox{D0}\rangle\!\rangle_U =
\sum_{p} \exp\left(\sum_{n\geq 1}  {1\over n} \alpha^i_{-n}
\bar\alpha^i_{-n}\right)|(p,p)\rangle\,.}
Let us expand out these boundary states, and in particular, consider
the contributions for $h=\bar{h}=0,1,2$. At $h=\bar{h}=0$,
both boundary states are proportional to the vacuum. At $h=\bar{h}=1$,
the $|\!|\hbox{D24}\rangle\!\rangle_U$ boundary state is proportional
to
\eqn\dtfeone{
- \sum_i \alpha^i_{-1} \bar\alpha^i_{-1}|(0,0)\rangle\,,}
while the $|\!|\hbox{D0}\rangle\!\rangle_U$ state is proportional to
\eqn\dzeroeone{
\sum_i \alpha^i_{-1} \bar\alpha^i_{-1}|(0,0)\rangle\,.}
The sum
$|\!|\hbox{D24}\rangle\!\rangle_U+|\!|\hbox{D0}\rangle\!\rangle_U$
therefore does not have any contribution at $h=\bar{h}=1$. This is
in agreement with the fact that the Monster theory does not have any
states of $h=\bar{h}=1$. At $h=\bar{h}=2$ we get for
$|\!|\hbox{D24}\rangle\!\rangle_U$
\eqn\tfetwo{
- {1\over 2} \sum_i \alpha^i_{-2}\bar\alpha^i_{-2}|(0,0)\rangle
+ {1\over 2} \sum_{i,j} \alpha^i_{-1}\alpha^j_{-1}
                  \bar\alpha^i_{-1} \bar\alpha^j_{-1} |(0,0)\rangle
+ \sum_{p:p^2=4} |(p,-p)\rangle \,,}
while for $|\!|\hbox{D0}\rangle\!\rangle_U$ we have
\eqn\dzeroetwo{
{1\over 2} \sum_i \alpha^i_{-2} \bar\alpha^i_{-2}|(0,0)\rangle
+ {1\over 2} \sum_{i,j}\alpha^i_{-1}\alpha^j_{-1}
                  \bar\alpha^i_{-1} \bar\alpha^j_{-1} |(0,0)\rangle
+ \sum_{p:p^2=4} |(p,p)\rangle \,.}
At $h=\bar{h}=2$ the sum of
$|\!|\hbox{D24}\rangle\!\rangle_U+|\!|\hbox{D0}\rangle\!\rangle_U$
therefore has the contribution
\eqn\dtotaltwo{
\sum_{i,j} \alpha^i_{-1}\alpha^j_{-1}
                  \bar\alpha^i_{-1} \bar\alpha^j_{-1} |(0,0)\rangle
+ \sum_{p:p^2=4}
{1\over \sqrt{2}} \bigl(|p\rangle + |-p\rangle\bigr)_L \otimes
{1\over \sqrt{2}}\bigl(|p\rangle + |-p\rangle\bigr)_R
\,.}
In the last sum we have written the momenta as tensor products of
left- and right-moving momenta.

Next we recall that the chiral Monster theory has $196884$ states
with $h=2$; of these there are ${24\cdot 25\over 2}$ states of
the form $\alpha^i_{-1}\alpha^j_{-1}|0\rangle$ and $98280$ states of
the form ${1\over \sqrt{2}} (|p\rangle+|-p\rangle)$ with $p^2=4$ (as
well as $24\cdot 2^{12}$ states coming from the twisted sector). What
the above calculation shows is that of the $196883^2$ $\W$-Ishibashi
states at $h=\bar{h}=2$, those that come from the untwisted sector
(\ie\ the $\left({24\cdot 25\over 2} + 98280 - 1\right)^2$ Ishibashi
states coming  from the first two types of states minus the
stress-energy tensor) contribute only if the left-label of the Monster
representation is the same as the right-label of the Monster
representation. Furthermore, all these diagonal states appear with the
same coefficient. If the D0-D24 combination is one of the boundary
states we have constructed before, then the group element $g$ must
therefore have the property that 
$D^{{\bf 196883}}_{ij}(g)=\delta_{ij}$ if 
$i$ and $j$ are untwisted labels. It is also clear that
$D^{{\bf 196883}}_{ij}(g)=0$ if $i$ and $j$ describe one untwisted and
one twisted label since the twisted sector of the geometric $\Zop_2$ 
orbifold (in which theory the D0-D24 boundary state is constructed)
consists of those states that are twisted with respect to both the
left- and the right- asymmetric orbifold. Thus the 
corresponding group element $g$ must give rise to a representation
matrix of the form
\eqn\twistone{
D^{{\bf 196883}}(g) = \pmatrix{ {\bf 1} & 0 \cr 0 & R(g)} \,,
}
where we have written the matrix in block-diagonal form, with the two
blocks corresponding to the untwisted and the twisted sector states,
respectively. Here $R(g)$ is a $98304\times 98304$ matrix, describing
the components of $g$ in the representation ${\bf 196883}$ with
respect to the twisted sector states.

\noindent We also know that, in the above notation,
\eqn\twisttwo{
D^{{\bf 196883}}(i) = \pmatrix{ {\bf 1} & 0 \cr 0 & -{\bf 1}} \,,
}
and thus it follows that
\eqn\twistthree{
D^{{\bf 196883}}( g i g^{-1} i ) = {\bf 1} \,.
}
Since the Monster group does not have any non-trivial normal
subgroups, it follows that $g i g^{-1} i = e$, and therefore that $g$
is in the centraliser of $i$, \ie\ in $C$. On the other hand, we know
how the elements in $C$ act on the $98579$ untwisted sector states of
${\bf 196883}$, and it is easy to see that there are only two elements
in $C$ that give rise to a matrix of the form \twistone\ above: either
$g=e$ or $g=i$. Let us choose to identify the fractional D0-D24
discussed above with  $|\!| e\rangle\!\rangle$. As we will see in the
next subsection, $|\!| i\rangle\!\rangle$ then corresponds to another
fractional D0-D24 at the  origin, differing from 
$|\!| e\rangle\!\rangle$ in the overall sign of the twisted sector
components of its boundary state. 

Apart from the direct boundary state argument given above, there
is another way to argue for the identification of $|\!|
e\rangle\!\rangle$ (and $|\!| i\rangle\!\rangle$) with a
fractional D0-D24 at the origin. We expect that such a D0-D24 is
at least invariant under the left-right symmetric (geometric)
action of a suitable lift of the simple Conway group $(\cdot 1)$
to the extra-special extension $C$ (since the elements in $(\cdot
1)$ correspond to Leech lattice automorphisms, which leave the
origin invariant). Indeed, it is obvious from \leftright\ that
both $|\!| e \rangle\!\rangle$ and $|\!| i \rangle\!\rangle$ are
in fact invariant under the left-right symmetric (diagonal) action
of any element of $C$. 
Furthermore, $e$ and $i$ are the only group elements in $\Mon$ with
this property.

\subsec{More fractional branes}

In subsection~3.2 we saw that by acting on the fractional D0-D24 brane
of subsection~3.1 with elements of the group $C=2_+^{1+24}(\cdot 1)$ acting
on the left, one obtains other fractional D-branes with a clear
geometric interpretation. From the previous section, we know that the
fractional D0-D24 is described by the boundary state
$|\!| e \rangle\!\rangle$, and in subsection~4.3 we showed that
the left action of an element $g$ of $C$ results in the boundary
state $|\!| g \rangle\!\rangle$. Thus the boundary states labelled
by an element of $C=2_+^{1+24}(\cdot 1)$ have a geometric
interpretation as fractional branes.

In particular, the boundary state $|\!| i \rangle\!\rangle$ associated
to the  involution $i$ corresponds to a fractional D0-D24 brane at the
origin, which differs from $|\!| e \rangle\!\rangle$ by the sign of
the twisted sector  components of its boundary state: if we normalised
the various boundary states so that $|\!| e\rangle\!\rangle$ is given
by \bsDzero, then
\eqn\bsi{
 |\!| i \rangle\!\rangle =
{1\over\sqrt{2}}\Bigl(|\!|\hbox{D0}\rangle\!\rangle_U
- |\!|\hbox{D0}\rangle\!\rangle_T
+ |\!|\hbox{D24}\rangle\!\rangle_U
- |\!|\hbox{D24}\rangle\!\rangle_T\Bigr)\,,}
where the subscripts $U$ and $T$ denote again the components in the
untwisted sector and in the sector twisted by the geometric $\Zop_2$,
respectively.

\subsec{Bulk branes}

None of the branes we described in section~4 have any moduli.
Indeed, as we saw in subsection~4.3, the self-overlap of any of
these branes leads to the one-loop partition function
$F(\tilde{q})=j(\tilde{q})-744$ in the open string, which
therefore does not have any massless
states.\footnote{$^\ddagger$}{Incidentally, this property of the
branes is special to the Monster theory; all other known self-dual
conformal field theories contain massless states in their chiral
partition function.} On the other hand, we saw in section~3.3 that
the theory should have a continuum of bulk D-brane states. It
therefore follows that bulk branes generically cannot preserve
$\W$. They are therefore examples of physical D-branes (preserving
the conformal symmetry) that are not captured by the construction
in section~4 (where we restricted our attention to D-branes
preserving the larger algebra $\W$).

One may wonder whether these bulk branes can be thought of as being
built out of fractional branes. In particular, one may expect
that combinations of fractional branes with a vanishing twisted sector
contribution can combine to form a bulk brane. The simplest example of
such a combination of fractional branes is described by the
superposition of $|\!| e \rangle\!\rangle$ and $|\!| i\rangle\!\rangle$.
In order to check whether this configuration of branes possesses
massless modes that describe the corresponding moduli, we determine
the cylinder diagram
\eqn\cyl{
\Bigl(\langle\!\langle e |\!|+\langle\!\langle i |\!|\Bigr)
\, q^{\half(L_0+\bar{L}_0-{c\over12})}\,
\Bigl(|\!| e \rangle\!\rangle+|\!| i \rangle\!\rangle\Bigr)
=2\chi_e(q)+2\chi_i(q)\,.
}
We want to write this amplitude in the open string channel, \ie\
in terms of the open string variable $\tilde q=\exp(-2\pi i/\tau)$. As
we have explained before,
\eqn\chie{
\chi_e(q)=j(\tau)-744=j(-1/\tau)-744
=\tilde q^{-1}+0+196884\tilde q+\cdots\,,
}
and thus we do not get any massless modes from $\chi_e$. (This is as
expected since $|\!| e \rangle\!\rangle$ and $|\!| i \rangle\!\rangle$
separately do not have any massless modes.) On the other hand,
$i$ lies in the class $2B$ of the ATLAS \refs{\atlas}, and using
\refs{\moon}, we find for the second term
\eqn\chii{
\chi_i(q)={\eta(\tau)^{24} \over \eta(2\tau)^{24}} + 24
=24+2^{12}\tilde q^{1/2}+\cdots\,.
}
In particular, the combined system has $48$ massless modes; $24$ of
these correspond to the moduli that describe the $24$ different
directions in which the bulk D0-brane can move off the fixed
point. (The other $24$ massless states correspond on the covering
space to an open string connecting a D0 and its image D0, which has
massless modes when the bulk brane is at a fixed point.)

Finally, let us remark that these `bulk branes' (which do not
preserve the full $\W$-symmetry) have higher mass than the
$\W$-preserving branes labelled by group elements in $\Mon$ (as
follows by comparing their boundary states in the limit
$q\rightarrow 0$ \refs{\polchinski,\tensiondimension}). It
therefore seems plausible that the lightest branes of the theory
preserve the full $\W$-symmetry, and therefore that they fall into
a representation of the Bimonster group.

\newsec{Conclusions}

In this paper we have shown that the D-branes preserving the
chiral algebra of Monster invariants transform in the regular
representation of both copies of the Monster group (and define a
representation of the Bimonster). Although this does not give a
complete classification of all possible D-brane states in the
Monster theory, it does provide evidence that the Bimonster
symmetry of the perturbative spectrum extends to a nonperturbative
symmetry of the full theory. In particular, it seems likely that
the D-brane states we have constructed are the lightest D-branes
in the spectrum, and that all other D-brane states can be formed
as composites of these building blocks.

In this paper we have restricted our attention to conformal field
theories that only consist of a vacuum sector, \ie\ that are
self-dual. This assumption guaranteed that the decomposition \decom\
(that is only known to hold for the vacuum sector \refs{\DLMimrn}) can
be used to decompose the full space of states as in \fulldecom. It
seems plausible that the decomposition \decom\ may hold more generally
for an arbitrary representation. This would then suggest that our
construction may generalise further. This idea is also supported by
the observation that our result is structurally very similar to what
was found in \refs{\GRW} for the (non-self-dual) WZW model
corresponding to $su(2)$ at level $k=1$. We hope to come back to this
point in a future publication.

The techniques described here may also be useful in trying to
obtain a more systematic understanding of D-branes in asymmetric
orbifolds. (Previous attempts at constructing D-branes in
asymmetric orbifolds have been made in \refs{\brr,\kors}.) In
particular, some of the branes described above (namely those that
correspond to group elements  in $\Mon \setminus C$) actually
involve Ishibashi states from asymmetrically twisted closed string
sectors.

Finally, we hope that the perspective we have discussed here might
be of use to mathematicians trying to obtain a more conceptual
understanding of monstrous moonshine. Conformal field theories with
boundaries have been less well developed in the
mathematical literature. In a physical framework the boundaries are
associated to D-branes and open strings which have endpoints on the
D-brane.  We have shown here that the McKay-Thompson series which are
the subject of the genus zero moonshine conjectures arise naturally in
the open string sector of the closed string theory with Monster
symmetry.  Perhaps this will suggest new approaches to the moonshine
conjectures.

\vskip 1cm

\centerline{{\bf Acknowledgements}}\pano

We would like to thank Terry Gannon, Peter Goddard, Emil Martinec,
and Andreas Recknagel
for useful conversations; and Richard Borcherds, and John McKay
for helpful communications.

The work of BC is supported by DOE grant DE-FG02-90ER-40560 and NSF
grant PHY-9901194.
MRG is grateful to the Royal Society for a University
Research Fellowship. He also acknowledges partial support from the EU
network `Superstrings' (HPRN-CT-2000-00122), as well as from the PPARC
special grant `String Theory and Realistic Field Theory',
PPA/G/S/1998/0061. The work of JH is supported by NSF grant PHY-9901194.

\listrefs

\bye